\documentclass[useAMS,usenatbib]{mn2e} 
\usepackage{graphicx} 
\usepackage{epstopdf} 
\usepackage{amsmath} 
\usepackage{amssymb} 
\title[]{Analytical Solutions of Radiative Transfer Equations in Accretion Discs with Finite Optical Depth} 
\author[ M. Samadi, F. Habibi, S. Abbassi]{ 
M. Samadi$^{1,3}$\thanks{samadimojarad@um.ac.ir},F. Habibi $^{2}$\thanks{f\_habibi@birjand.ac.ir}, S. Abbassi $^{1}$\thanks{abbassi@um.ac.ir} 
\\ 
$^{1}$Department of Physics, Faculty of Sciences, Ferdowsi University of Mashhad, Mashhad, 91775-1436, Iran\\ 
$^{2}$Department of Physics, Faculty of Sciences, University of Birjand, Birjand, Iran\\ 
$^{3}$Research Institute for Astronomy and Astrophysics of Maragha (RIAAM)- Maragha, P. O. Box: 55134 - 441, Iran} 
\date{} 
\begin{document} 
\pagerange{\pageref{firstpage}--\pageref{lastpage}} \pubyear{2020} 
\maketitle \label{firstpage} 
\begin{abstract} 
The main purpose of this paper is to obtain analytical solutions for radiative transfer equations related to the vertical structure of accretion discs with finite optical depth. In the non-gray atmosphere, we employ the optical-depth dependent Eddington factor to define the relationship between the mean intensity and radiation stress tensor. Analytical solutions are achieved for two cases: (i) radiative equilibrium, and (ii) a disc with uniform internal heating and both cases are assumed to be in local thermodynamical equilibrium (LTE), too. These solutions enable us to study probable role of scattering and disc optical depth on the emergent intensity and other radiative quantities. Our results show that for the first case, the surface value of mean intensity with constant Eddington factor is three times larger than that with variable factor. Moreover, scattering has no role in the vertical radiative structure of discs with the assumptions of the first case.
On the other hand, for the second case, we encounter reductions in all radiative quantities as the photon destruction probability decreases (which is equivalent to increasing scattering). Furthermore, for both cases with total optical depth less than unity, the outward intensity towards the polar direction becomes less than that from the edges of disc which is contrary to limb-darkening. At the end, we apply our results to find the spectrum from accretion systems, based on two dynamical models. Consequently, we can see that how the total optical depth varies with frequency and causes remarkable changes on the emergent spectra.
\end{abstract} 
\begin{keywords} 
accretion, accretion discs, black hole physics, opacity, radiative transfer, scattering, methods: analytical 
\end{keywords} 
\section{INTRODUCTION} 
Since basic theoretical models of accretion discs have been proposed to explain pivotal features of emergent spectrum arising from medium around black holes, there are still some difficulties to work on the vertical structure of such systems. The standard model of accretion discs (Shakura \& Sunnyev 1973, Novikov \& Thorne 1973, Lynden-Bell \& Pringle 1974) provides detailed formulas to find radial dependency of physical quantities. One of these quantities is the local effective temperature that substitution of this temperature in the blackbody flux equation yields the theoretical emergent spectrum. 
Nevertheless, the radiation field affects significantly the local properties (such as the local temperature) of the regions where it passes. Due to the possibilities of emission, absorption and scattering, the behavior of radiation field becomes very complex to study. Generally, radiation and matter are often coupled to each other and hence more precise solutions of the vertical structure needs to consider the radiative transfer (and its moments) beside hydrodynamic equations. Since the absorption, emission and scattering depend on the frequency of photons, solving the complete set of dynamical and radiative equations sounds impossible without some simplifying assumptions like gray atmosphere or local thermodynamic equilibrium. 
Unlike the stellar atmosphere model, the proposed accretion disc models have been more ambiguous about radiation aspects and they have not been adequate to match the whole of observed spectra, yet (Wang et al. 1999). Although, there are serious differences between stellar atmosphere and accreting mediums in optical depth, source of energy, gravitational acceleration and scattering types (Adam et al. 1988), they are still convenient patterns to establish sophisticated solutions of radiation transfer and description of emission spectrum from the vicinity of black holes. Subsequently, some numerical (by simulations: Park 1993, Curd \& Narayan 2018, Ryan et al. 2018 or especial methods, like Monte Carlo: Boissé 1990, Dolence et al. 2009, Foucart 2018, or discrete ordinates: Kanschat 1990, Stenholm 1991) and theoritical (Hubeny 1990, Cao et al. 1998, Burigana 1995, Baschek et al. 1997, Danielian 2010, Jankovic et al. 2018) researches have been done to study statistic and moving mediums with thick or thin optical depth. Furthermore, several authors have achieved analytical solutions based on common assumptions (Kalkofen\& Wehrse 1982, Hubeny 1990, Kryzhevoi et al. 2001, , Fukue \& Akizuki 2006, Boss 2009). Other elaborate works have been done in the relativistic regime to study the radiative transfer problem in relativistically moving media such as Anderson \& Spiegel (1972), Udey \& Israel (1982), Thorne (1981), Zane et al. (1996), Younsi et al. (2012), Takahashi \& Umemura (2017) and Fukue (2014), (2017) and (2018). 
As we know, the emission spectrum from inner parts of X-ray binaries is affected by mostly electron scattering opacity. Shakura \& Sunyaev (1973) took into account this opacity and believed that it has the main role in some regions of accretion discs around black holes. When Thomson scattering is dominant, its notable effect can be seen in a decrease of the observed spectrum (modified blackbody spectrum) by the factor of $\sqrt{\epsilon_\nu}$ ($\epsilon_\nu=\sigma_\nu/(\sigma_\nu+\kappa_{abs})$ where $\sigma_\nu$ is the opacity coefficient for scattering and $\kappa_{abs}$ is the absorption opacity and $\epsilon_\nu$ is called the photon destruction probability) comparing with the blackbody spectrum. In fact, this flatted spectrum is called as the modified blackbody spectrum and it leads us to consider a frequency-dependent correction for elastic scattering (Czerny \& Elvis 1987). 
Many researches have been done to calculate the emergent spectrum from accreting systems based on the previous available knowledge from stellar atmospheres and the assumption of unique effective temperature. So the work of Laor \& Netzer (1989) might have absorbed more attentions because they took into account a vertical temperature gradient in order to calculate spectra from massive thin accretion discs. They found out that the effective and surface temperatures are approximately equal. Considering non-isotermal atmosphere led them to conclude that the electron scattering is not as noticeable as one in other models with higher temperatures. Other approach for determination of emergent spectrum has been introduced by Wang et al. (1999). They applied the height-averaged equations but they calculated the spectrum of a slim disc based on its radial structure. In both mentioned works, the internal heating were not involved and also the optical depth was considered as infinity. Fukue (2011, hereafter Fu11) and (2012) studied scattering effects on the emergent intensity and other radiative quantities. For several certain cases, he could solve analytically the radiative transfer equations with finite optical depth. The result of his work showed that the source functions are smaller than the thermal spectra when the effect of finite optical depth and scattering are combined and hence they were provided another form of a modified blackbody, proportional to $\epsilon_\nu$ instead of $\sqrt{\epsilon_\nu}$. 
In this paper, we follow Fu11 and consider geometrically thin accretion disc with finite optical depth for two different cases. 
To simplify equations we need to employ Eddington factor which provides a certain relationship between mean intensity and radiation stress.The Eddington factor that we will use in this paper is a function of the optical depth (Tamazawa et al. 1975).
We will concentrate on scattering and disc optical depth effects on the radiative transfer, mean intensity, the Eddington flux and the mean radiation stress. 
The outline of this paper is as follows. In Section 2, we present the basic radiation equations, which include the moments of the transfer equation. In the following, we work on these cases: firstly (in \S2) radiative equilibrium and secondly (\S4) a flow with uniform heating. 
In section 5, we find the frequency dependency of radiative quantities for several accretion systems. And finally, we summarize our whole results in section 6. 

\section{Basic Assumptions and Equations}\label{S2} 
Generally, the radiative transfer equation has to be solved in four dimensional space-time, but it will be too difficult to obtain any solutions. Hence, in order to solve radiative transfer equations of accretion discs, we need some basic assumptions. Firstly we assume that the disc is static in the corotating frame. Secondly, it is assumed to be geometrically thin and locally plane parallel. Moreover, we suppose that convection and conduction are ignorable to transport the energy in the vertical direction and hence just the radiation exist to move energy outwards. Consequently, the vertical disc's atmosphere would be well approximated by one dimensional equations. In this paper, we adopt the non-gray approximation which implies that the opacity depends on frequency. With the assumptions stated above, we can analyze the behavior such a system by radiative transfer and hydrodynamic equations in the vertical direction. First of all we refer to the three basic equations: the frequency-dependent transfer equation and zeroth and first moments of it (Mihalas 1978, Rybicki \& Lightman 1979, Mihalas \& Mihalas 1984, Shu 1991, Kato et al. 2008) which are simplified by applying the mentioned assumptions as: 
\begin{equation} 
\label{phiL} 
\mu \frac{dI_\nu}{dz}=\rho\bigg[\frac{j_\nu}{4 \pi}-(\kappa_\nu+\sigma_\nu)I_\nu +\sigma_\nu J_\nu\bigg], 
\end{equation} 
\begin{equation} 
\label{phiL1} 
\frac{dH_\nu}{dz}=\rho\bigg[\frac{j_\nu}{4 \pi}-\kappa_\nu J_\nu\bigg], 
\end{equation} 
\begin{equation} 
\label{phiL2} 
\frac{dK_\nu}{dz}=-\rho\big(\kappa_\nu+\sigma_\nu\big)H_\nu, 
\end{equation} 
where $I_\nu$, $J_\nu$, $H_\nu$ and $K_\nu$ are the specific intensity, the mean intensity, the Eddington flux and the mean radiation stress, respectively. Other factors and parameters are seen in these three equations: $j_\nu, \mu(=\cos\theta), \rho$ and $c$ show the mass emissivity, the direction cosine, the gas density and the light velocity, respectively. Under the non-gray treatment, all radiative quantities are variable and functions of frequency. The dependency of all parameters and radiative quantities to the frequency is clearly presented by index of $\nu$.\\ 
\begin{equation} 
\mu \frac{dI_\nu}{d\tau_\nu}=I_\nu -S_\nu, 
\end{equation} 
\begin{equation} 
\frac{dH_\nu}{d\tau_\nu}=J_\nu-S_\nu, 
\end{equation} 
\begin{equation} 
\frac{dK_\nu}{d\tau_\nu}=H_\nu, 
\end{equation} 
\begin{displaymath} 
S_\nu=\frac{1}{\kappa_\nu+\sigma_\nu}\frac{j_\nu}{4\pi}+\frac{\sigma_\nu}{\kappa_\nu+\sigma_\nu}J_\nu 
\end{displaymath} 
If we assume the accretion flow is in LTE, i.e. $j_\nu=4\pi\kappa_\nu B_\nu$ we will have, 
\begin{equation} 
S_\nu=\epsilon_\nu B_\nu+(1-\epsilon_\nu)J_\nu, 
\end{equation} 
As we mentioned before $\epsilon_\nu$ is the photon destruction probability and is a function of scattering and absorption opacity coefficients. 
In order to complete the system of equations, the vertical component of momentum equation and also energy equation for matter (Kato et al. 2008) are involved as: 
\begin{equation} 
\label{phik} 
-\frac{d\Phi}{dz}-\frac{1}{\rho}\frac{dp}{dz}+\frac{\kappa_\nu +\sigma_\nu}{c}4 \pi H_\nu=0, 
\end{equation} 
and 
\begin{equation} 
\label{phik1} 
q^+_{vis}-\rho \int\big(j_\nu -4 \pi \kappa_\nu J_\nu) d\nu=0, 
\end{equation} 
where $\Phi$ is the gravitational potential, $p$ the gas pressure, and $q^+_{vis}$ the viscous-heating rate. We have supposed that the density distribution would be adjusted so as to hold the hydrostatic equilibrium (Eq.8) through the main part of the disc atmosphere, under the radiative flux obtained later (Fukue 2006). Therefore, we do not solve Eq.(8). 
In the following, we define the optical depth $\tau_{\nu}$ as: $ 
d\tau_\nu \equiv -\rho \big(\kappa_\nu +\sigma_\nu \big)dz. 
$ 
Therefore, total optical depth of the disc becomes: 
\begin{equation} 
\tau_{\nu 0} =- \int_{h}^{0} \rho (\kappa_\nu +\sigma_\nu) dz, 
\end{equation} 
where $h$ is the half-thickness of disc. 
It is still hard to solve this system of equations with several kinds (i.e. algebraic, differential and integral equations), here we employ the Eddington approximation which provides a relationship between the mean values of radiation stress and intensity and consists of a model for opacity. The variable Eddington factor, $f_{Edd}$ is given by (Tamazawa et al. 1975): 
\begin{equation} 
f_{Edd}= \frac{1+\tau_\nu}{1+3\tau_\nu}
\end{equation} 
and this is used in the Eddington approximation, which is written as: 
\begin{equation} 
K_\nu =f _{Edd} J_\nu, 
\end{equation} 
\begin{figure} 
\centering 
\includegraphics[width=90mm]{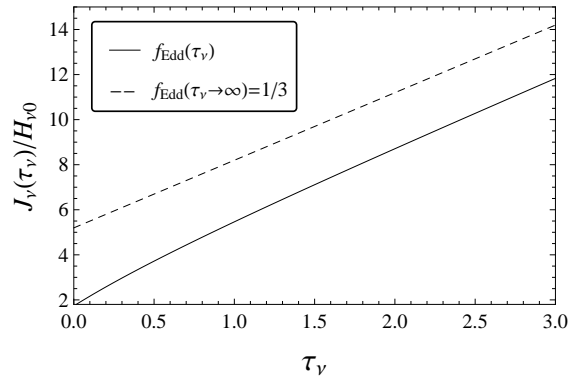} 
\caption{Solutions for the assumption of radiative equilibrium: Variation of mean intensity, $J_\nu$ normalized by $H_{\nu0}$ (which shows the Eddington flux at zero optical depth) 
with respect to the optical depth in the case of radiative equilibrium with two assumptions: 1. $f_{Edd}(\tau_\nu)$ and 2. constant Eddington factor, $f_{Edd}(\tau_\nu\rightarrow \infty)=1/3$. } 
\end{figure} 
In the next two sections, we solve analytically the basic equations of this system with using the Eddington factor and for two cases of (i) the radiative equilibrium (RE), and (ii) a flow having internal uniform heating. In order to compare our results with those in Fu11, we will display the constant Eddington factor $f_{Edd}=1/3$ as $f_0$ and the variable $f_{Edd}$ as $f_\nu$. 
\begin{figure*} 
\centering 
\includegraphics[width=180mm]{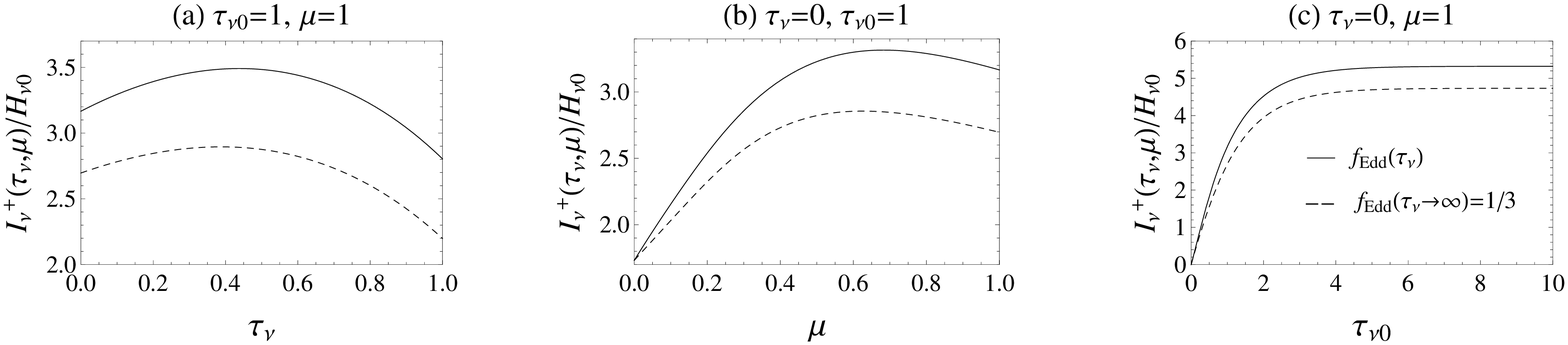} 
\caption{Solutions for the assumption of radiative equilibrium: Outward intensity (normalized by $H_{\nu0}$) as a function of (a) the optical depth, $\tau_\nu$, (b) direction cosine, $\mu$, and (c) disc optical depth, $\tau_{\nu0}$ with two assumptions 1. $f_{Edd}(\tau_\nu)$ and 2. constant Eddington factor $f_{Edd}(\tau_\nu\rightarrow \infty)=1/3$. }  
\end{figure*} 
\section{Radiative Equilibrium }\label{S5} 
If there are no sources of cooling or heating in the flow, the current situation is considered as radiative equilibrium and we will have: 
\begin{displaymath} 
j_\nu = 4 \pi \kappa_\nu J_\nu, 
\end{displaymath} 
with this assumption the energy equation (9) is satisfied easily. With the help of the Eddington approximation (Eq.12) and using Eq.(7) and (10), the radiative transfer equations (1)-(3) can rewrite as following: 
\begin{equation} 
\mu \frac{dI_\nu}{d\tau_\nu}=I_\nu -J_\nu,\label{f1} 
\end{equation} 
\begin{equation} 
\frac{dH_\nu}{d\tau_\nu}=0,\label{f2} 
\end{equation} 
\begin{equation} 
\frac{d(f_{Edd} J_\nu)}{d\tau_\nu}=H_\nu,\label{f3} 
\end{equation} 
The equation (14) shows that radiative flux $H_\nu$ does not depend on the optical depth and has a constant value $H_{\nu 0}$, 
\begin{equation} 
H_\nu=H_{\nu 0}.\label{f4} 
\end{equation} 
Now, the integral of equation (15) can be easily calculated to give a solution for mean intensity $J_\nu$ and the mean radiation stress $K_\nu$ as: 
\begin{displaymath} 
J_\nu=\frac{H_{\nu 0}\tau_\nu+J_{\nu 0}}{f_{Edd}}, 
\end{displaymath} 
\begin{equation} 
K_\nu=H_{\nu 0}\tau_\nu +K_{0}. 
\end{equation} 
where $J_{\nu 0}$ and $K_{0}$ are both integration constants. Here, we specify boundary conditions as: $J_{\nu 0}=H_{\nu 0} c_\nu$ at $\tau_{\nu 0}=0$, where $c_\nu= \sqrt{3}$ (Mihalas \& Mihalas 1984). 
Finally, we achieve the analytical solution for $J_\nu$ as: 
\begin{equation} 
\text{J}_\nu= H_{\nu 0}
\frac{\tau_\nu+c_\nu}{f_{Edd}}
\end{equation} 
By using constant Eddington factor (i.e. $f_{Edd}=f_0=1/3$), the mean intensity is a linear function of $\tau_\nu$.
In Figure (1), we have plotted the function of $J_\nu$ normalized by $H_{\nu0}$. As seen, the value of mean intensity becomes smaller with $f_{Edd}(\tau_\nu)$ and the difference between two $J_\nu$'s is noticeable at smaller optical depths. 
The reason of smaller value of the mean intensity with $f_\nu$ can be explained or understood from Eq.(11) and (12). According to these two equations, $J_\nu$ becomes maximum value, $J_{\nu max}$, and equal to $3K_\nu$ when $\tau_\nu$ tends to infinity. However, for any other optical depth less than infinity, we find $J_\nu<3K_\nu$. For instance, $J_\nu=2 K_\nu$ at the level with $\tau_\nu=1$. On the other hand, in the disc's surface ($\tau_\nu=0$) we have $J_\nu=B_\nu$ and this is the minimum possible value of the mean intensity. Therefore, we would overestimate $J_\nu$ and find $J_{\nu max}$ for all $\tau_\nu$'s, if we applied the constant Eddington factor, $f_0$ in the relationship between $K_\nu$ and $J_\nu$ just like Fu11 had used.

Now if we compare Eq.(5) with (14), we can find: 
\begin{displaymath} 
S_\nu- J_\nu=0, 
\end{displaymath} 
so the source function becomes equal to the mean intensity. In this case, we could find solutions without using the Planck function. Due to the assumption of LTE, $B_\nu$ is equal to $J_\nu$, too. Comparing two $J_\nu$'s and employing the relation of $\int^\infty_0B_\nu d\nu=\sigma T^4/\pi$, we can estimate the surface temperature, $T_s$ for two cases of $f_0$ and $f_\nu$ seperately as:
\begin{displaymath}
T_s=\bigg(\frac{\pi}{\sigma}H_{\nu0}\frac{c_\nu}{f_{Edd}|_{\tau_\nu=0}}\bigg)^{1/4}
\end{displaymath}
from this equation, we find two different values of $T_s$ for constant and variable Eddington factors:
\begin{displaymath}
T_s=\bigg(\frac{\pi}{\sigma} H_{\nu0}c_\nu\bigg)^{1/4}\times
\begin{cases}
3^{1/4}\hspace*{0.5cm}\text{if}\hspace*{0.2cm} f_{Edd}=f_0\\
1\hspace*{0.9cm}\text{if}\hspace*{0.2cm} f_{Edd}=f_\nu
\end{cases}
\end{displaymath}
then we have $T_s(f_0)/T_s(f_\nu)\approx1.32$ which declares that the surface temperature is estimated larger with the constant Eddington factor.


Using the solutions in Eq.(16)-(18), we can solve the differential transfer equation (13) and obtain the specific intensity as: 
\[ 
I_\nu^\pm(\tau_\nu,\mu)= H_{\nu 0} 
\bigg\{ 3( \tau_\nu \pm \mu+c_{\nu})-2\pm 
\frac{2}{\mu} (c_\nu-1)\times 
\] 
\begin{equation} 
\hspace*{1.5cm} e^{\pm(\tau_\nu+1)/\mu} Ei\big[\mp\frac{\tau_\nu+1 }{\mu}\big]\bigg\}+C_\pm e^{\pm\tau_\nu/\mu},\label{I1} 
\end{equation} 
where the sign of $\pm$ is positive for outward intensity, $I_\nu^+$, and it is negative for inward one, $I_\nu^-$. Notice that $C_\pm$ is integral constant ($C_-$ used for inward intensity and $C_+$ for outward one) and can be obtained with a proper boundary condition. The exponential integral function $Ei[\tau_\nu]$ is seen in Eq.(19) which is defined as $Ei[\tau_\nu]=-\int_{-\tau_\nu}^\infty e^{-t}/t dt$. In the following, we find the constants of $C_+$ and $C_-$. Firstly, we determine $C_-$ for the inward intensity, $I_\nu^-(\tau_{\nu},\mu)$ which is easily specified by reading negative sign of $\pm$ in equation (19): 
 \[
\text{I}_\nu^-(\tau_{\nu},\mu)=C_- e^{-\tau/\mu}-H_{\nu 0}
\bigg\{  2-3( c_{\nu} - \mu+\tau_\nu)
\]
\begin{equation} 
\hspace*{1cm}+2 \frac{c_\nu-1}{\mu} e^{-(\tau_\nu+1)/\mu} Ei\big[\frac{\tau_\nu+1}{\mu}\big]\bigg\}
\end{equation} 
In order to gain $C_-$, we have the following boundary condition: 
\begin{equation} 
I_\nu^-(0,\mu)=0, 
\end{equation} 
which is valid in the absence of irradiation. With applying this condition in Eq.(21) we will have: 
\begin{equation} 
\text{C}_-=H_{\nu 0}
\bigg\{ 3(\mu - c_{\nu}) +2+ \frac{2}{\mu}(c_{\nu}-1) e^{-1/\mu} Ei\big[\frac{1}{\mu}\big]\bigg\},
\end{equation} 
\begin{figure*} 
\centering 
\includegraphics[width=180mm]{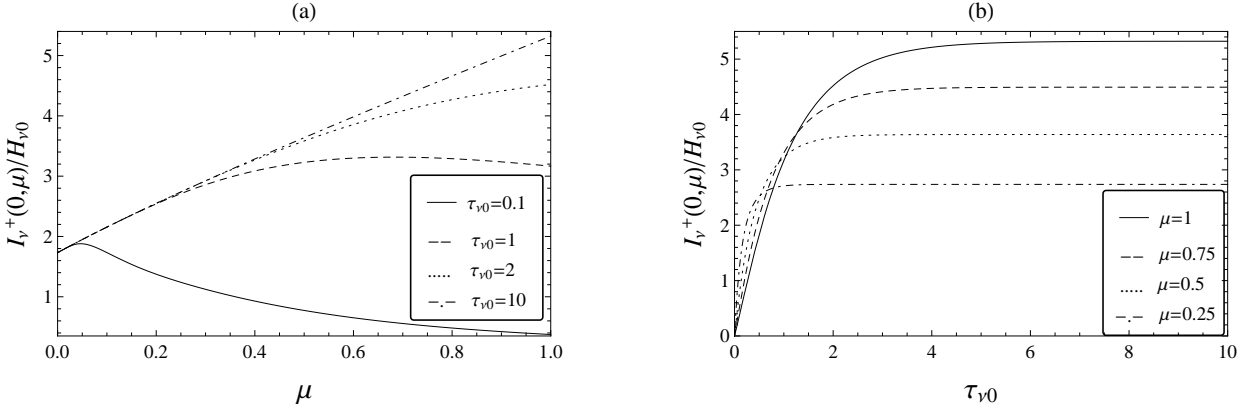} 
\caption{Solutions for the assumption of radiative equilibrium: (a) Emergent intensity (normalized by $H_{\nu0}$) as a function of the direction cosine, $\mu$, with four values of $\tau_{\nu0}$. (b) Variation of normalized emergent intensity with the disc optical depth, $\tau_{\nu0}$ and for four $\mu$'s. In this figure, we have assumed $f_{Edd}=f_\nu$.} 
\end{figure*} 
Now we can write the outward intensity, $I_\nu^+ (\tau_{\nu},\mu)$ from Eq.(19) as: 
\[
\text{I}_\nu^+(\tau_\nu,\mu)=C_+ e^{\tau_\nu/\mu}+ H_{\nu 0}
\bigg\{-2+3( \tau_\nu+ \mu+c_{\nu})\]
\begin{equation} 
\hspace*{1cm}+ 2\frac{c_\nu-1}{\mu} e^{(\tau_\nu+1)/\mu} Ei\big[-\frac{ \tau_\nu+1}{\mu}\big] \bigg\}\\ 
\end{equation} 
To determine the constant of $C_+$, we should use another boundary condition based on the assumption of finite optical depth. It is reasonable to suppose that $I_\nu^+ (\tau_{\nu 0},\mu)$ consists of two terms, one constant and the other one variable: 
\begin{equation} 
I_\nu^+(\tau_{\nu 0},\mu)=I_{\nu 0} + I_\nu^-(\tau_{\nu 0},\mu) 
\end{equation} 
where $I_{\nu 0}$ is the uniform incident intensity, showing the equatorial heating rate and the second term is the inward intensity from the back side of the flow beyond the midplane (Fukue \& Akizuki 2006, Fukue 2012). 
After some manipulations, we can calculate the constant of $C_+$ as: 
\[
\text{C}_+=C_- e^{-2\tau_{\nu0}/\mu}+e^{-\tau_{\nu0}/\mu}
\bigg\{I_{\nu0}-6\mu H_{\nu 0}-2(c_{\nu}-1)\times\]
\begin{equation} 
\hspace*{0.3cm}\frac{H_{\nu0}}{\mu}\bigg[e^{1/\mu}Ei\big[-\frac{\tau_{\nu0}+1}{\mu}\big]
+e^{-(2\tau_{\nu0}+1)/\mu}Ei\big[\frac{\tau_{\nu0}+1}{\mu}\big]\bigg]\bigg\}
\end{equation} 
It is also important to determine the emergent intensity, $I_\nu^+(\tau_\nu=0,\mu)$, which comes out from the disc's surface: 
\begin{displaymath} 
\text{I}_\nu^+(0,\mu)=C_++H_{\nu 0}
\big\{ 3(\mu+ c_{\nu})-2+ 
2\frac{c_\nu-1}{\mu} e^{1/\mu} Ei\big[-\frac{1}{\mu}\big]\big\} 
\end{displaymath} 
Notice that here just the constant $C_+$ depends on the disc optical depth and we can determine it for very small ($\tau_{\nu0}\rightarrow 0$) and very large ($\tau_{\nu0}\rightarrow \infty$) optical depths of the disc which are referred to optically thin and optically thick discs, respectively. 
At first, for $\tau_{\nu0}\rightarrow 0$ we have

\begin{displaymath} 
e^{-\tau_{\nu0}/\mu}\sim 1-\frac{\tau_{\nu0}}{\mu} 
\end{displaymath} 

Here we use the expansion of exponential function and just keep two first terms of it in the limit of $x\rightarrow 0$, 
\begin{displaymath} 
e^x=\Sigma \frac{x^n}{n!}=1+x+\frac{x^2}{2!}+...
\end{displaymath}

for other terms including exponential function, we do the same and keep just two first terms. On the other hand, for $\tau_{\nu0}\rightarrow \infty$ we find out that $C_+\rightarrow 0$ and hence the emergent intensity becomes independent of the disc optical depth (cf. Fig.3b) and has formed as a linear function of $\mu$ with $f_{Edd}=f_0$ (and approximately with $f_{Edd}=f_\nu$ too, according to Fig.3a). 
In figures (2)-(4), we see variation of the outward intensity with respect to (i) the optical depth, $\tau_\nu$, (ii) the direction cosine, $\mu$ and (iii) the disc optical depth, $\tau_{\nu0}$. Fig.(2) shows the differences between solutions with two assumptions for the Eddington factor. As seen, $I_\nu^+$ is greater with $f_{Edd}=f_\nu$ in all three panels of Fig.(2) (this is due to smaller $J_\nu$ with negative sign in Eq.13). The effect of $\mu$ and $\tau_{\nu0}$ have been examined for the case of $f_{Edd}=f_\nu$ in Fig.(3). The first panel of this figure reveals that with $\tau_{\nu0}\geq 1$ the outward intensity increases as the optical depth increases and this causes the familiar effect of limb-darkening (cf. Fig.3 of Fukue 2012). However, in discs with very small optical depth, means $\tau_{\nu0}<1$, something different happens and we see the edge of disc brighter than its center even when our line of sight is aligned with the disc's polar axis (Fukue \& Akizuki 2006). Moreover, according to Fig.(3b), with larger values of $\tau_{\nu0}(\gtrsim 2)$, $I_\nu^+$ approximately remains constant in all directions, means that disc looks like optically thick even with a limited optical thickness. 

Figure (4) displays the behavior of the outward intensity with respect to the optical depth. With $\tau_{\nu0}\geq 1$ and for all directions, we see that $I_\nu^+$ grows gradually with increasing $\tau_\nu$ and in some directions it reaches a peak at a certain optical depth. As seen in panel (a), the outward specific intensity reduces towards inside the disc but just in a translucent disc with $\tau_{\nu0}=0.1$ and in a direction not so close to the vertical axis (where $\mu\rightarrow 0$). Moreover, in all discs with any optical depth, the outward intensity varies almost linearly with respect to $\tau_\nu$ from the vertical direction ($\mu\simeq 0$). In the next section, we will see similar trends in all radiative quantities. In addition, we will find their dependency to the scattering too which was absent here. 
\begin{figure*} 
\centering 
\includegraphics[width=180mm]{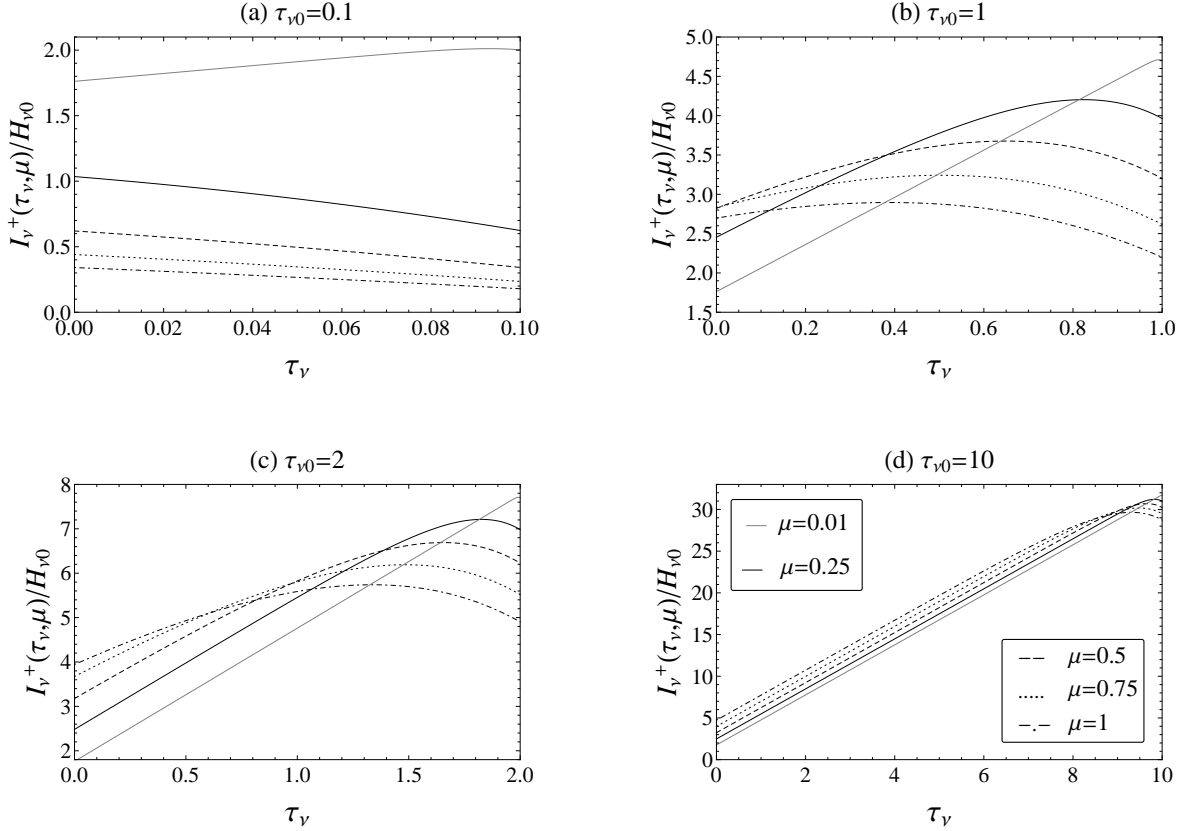} 
\caption{Solutions for the assumption of radiative equilibrium: Variation of outward intensity (normalized by $H_{\nu0}$) with respect to the optical depth. We have examined several values of the direction cosine, $\mu$, in each panel in a disc with certain total optical depth. In these plots, we have assumed $f_{Edd}=f_\nu$. } 
\end{figure*} 
\section{Uniform heating} 
For the second case, we suppose that the current heating inside the flow does not depend on the optical depth which is here called uniform heating. In this section, we also employ the local thermodynamic equilibrium $j_\nu=4\pi\kappa_\nu B_\nu$ in Eq.(9) and find: 
\begin{equation} 
q^+_{vis}=4\pi\rho\int\kappa_\nu(B_\nu-J_\nu)d\nu, 
\end{equation} 
Composing Eq.(5) and (7) gives us: 
\begin{equation} 
\frac{dH_\nu}{d\tau_\nu}=-\epsilon_\nu(B_\nu-J_\nu), 
\end{equation} 
Comparing these two equations, we see that the right-hand side of Eq.(27) is similar to the heating term. Therefore, the uniform heating assumption provides a constant term in the right-hand side of Eq.(27): 
\begin{equation} 
\epsilon_\nu(B_\nu-J_\nu)=q_\nu, 
\end{equation} 
Now we can easily integrate Eq.(27) and obtain: 
\begin{equation} 
H_\nu(\tau_\nu)=H_{\nu0}\bigg(1-\frac{\tau_{\nu}}{\tau_{\nu0}}\bigg), 
\end{equation} 
where we have assumed that $H_\nu(0)=H_{\nu0}$ and $q_\nu=H_{\nu0}/\tau_{\nu0}$. To determine $K_\nu$, we substitute Eq.(28) in Eq.(6) and integrate it, we will have: 
\begin{equation} 
K_\nu = H_{\nu0}\bigg( 1-\frac{\tau_\nu}{2 \tau_{\nu 0}}\bigg)\tau_\nu+K_{\nu0}, 
\end{equation} 
where $K_\nu(0)=K_{\nu0}=f_{Edd}J_{\nu0}$. 

In figure (5), we have plotted the Eddington flux and mean radiation stress via the optical depth. As it can be seen, $H_\nu$ decreases linearly with increasing $\tau_\nu$ for all values of $\tau_{\nu0}$ and $\epsilon_\nu$. It might seem unclear that how $H_\nu$ depends on $\epsilon_\nu$ whereas it is not seen any dependency to $\epsilon_\nu$ in Eq.(29). This point will be explained after equation (34). Panels (d)-(f) reveals that the behavior of $K_\nu$ with respect to $\tau_\nu$ is not similar in discs with different optical depths. 

To determine $J_\nu$, we use Eq.(12) and employ the relation of $K_\nu$ from Eq.(30) we find: 
\begin{equation} 
J_\nu = \frac{H_{\nu0}}{f_{Edd}}\bigg( 1-\frac{\tau_\nu}{2 \tau_{\nu 0}}\bigg)\tau_\nu+J_{\nu0}, 
\end{equation} 
where $J_\nu(0)=J_{\nu0}=H_{\nu0}c_\nu$. 
From Eq.(7) and (28), we find $S_\nu=J_\nu+q_\nu$. Therefore, the source and Planck functions are achieved as: 
\begin{equation} 
S_\nu =H_{\nu0}\bigg[ \frac{1}{f_{Edd}}\bigg( 1-\frac{\tau_\nu}{2 \tau_{\nu 0}}\bigg)\tau_\nu+\frac{1+c_\nu\tau_{\nu 0}}{\tau_{\nu 0}}\bigg], 
\end{equation} 
\begin{figure*} 
\centering 
\includegraphics[width=180mm]{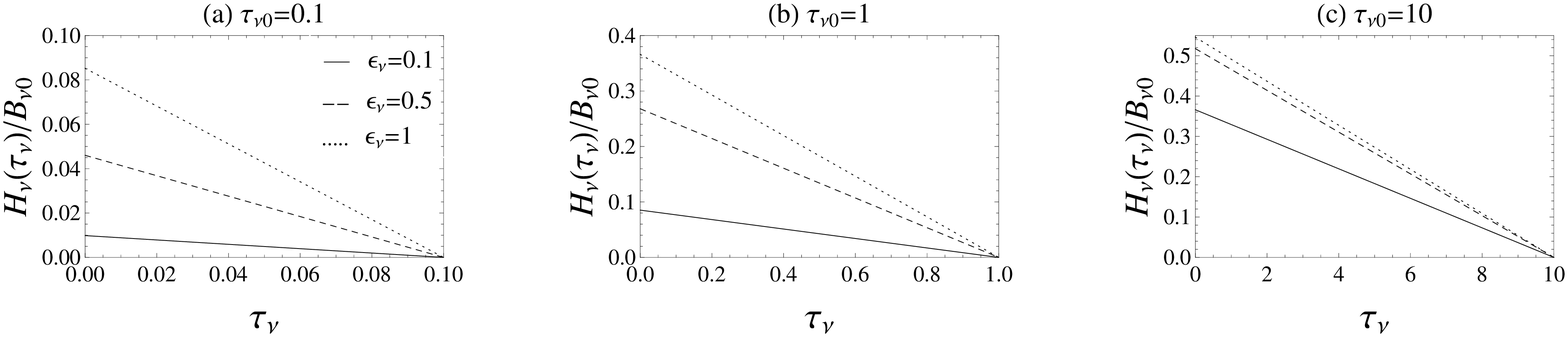} 
\includegraphics[width=180mm]{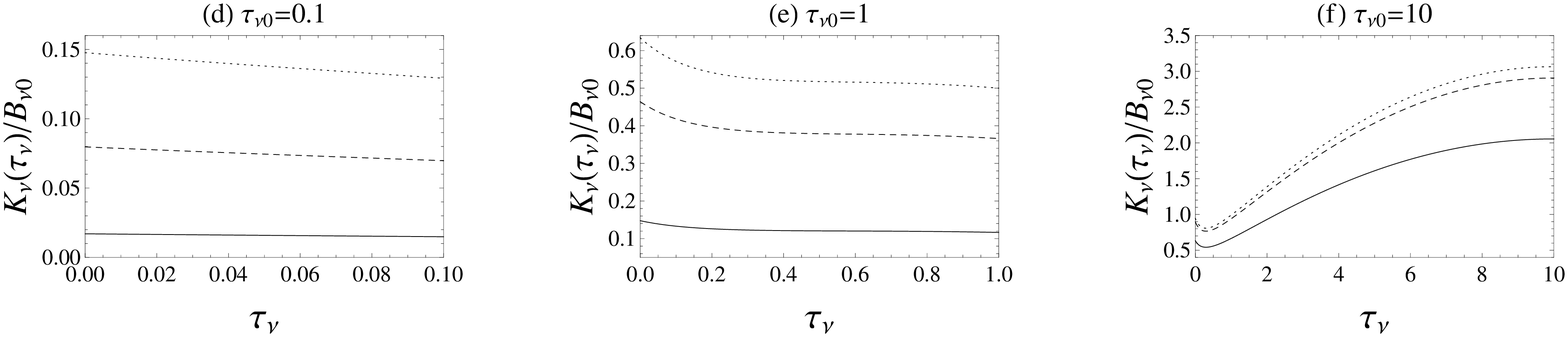} 
\caption{Solutions for uniform heating case: Variation of Eddington flux, $H_\nu$, and mean radiation stress, $K_\nu$, [both normalized by $B_{\nu0}=B_\nu(\tau_\nu=0)$] with respect to the optical depth, $\tau_\nu$. These solutions are obtained independently and without considering Eddington factor.} 
\end{figure*}  
And from Eq.(28) $B_\nu=J_\nu+q_\nu/\epsilon_\nu$ and then substituting Eq.(31) and the relation of $q_\nu=H_{\nu0}/\tau_{\nu0}$ in it, we find: 
\begin{equation} 
B_\nu =H_{\nu0}\bigg[ \frac{1}{f_{Edd}}\bigg( 1-\frac{\tau_\nu}{2 \tau_{\nu 0}}\bigg)\tau_\nu+\frac{1+c_\nu\epsilon_\nu\tau_{\nu 0}}{\epsilon_\nu\tau_{\nu 0}}\bigg], 
\end{equation} 
So we can determine the value of $H_{\nu0}$ with respect to $B_\nu(0)=B_{\nu0}$: 
\begin{equation} 
H_{\nu0}=B_{\nu0}\bigg[\frac{\epsilon_\nu\tau_{\nu 0}}{1+c_\nu\epsilon_\nu\tau_{\nu 0}}\bigg]. 
\end{equation} 
This relation enable us to study the effect of $\epsilon_\nu$ and therefore scattering influence on the radiative quantities. Without employing Eq.(34), we cannot find directly the dependency of scattering in all radiative equations exception for Eq.(33), hence we will substitute $B_{\nu0}$ in those equations instead of $H_{\nu0}$. 

Using Eq.(34) in (32), we find the source function at the disc's surface as:
\[S_\nu(\tau_\nu=0) =\frac{1+c_\nu\tau_{\nu 0}}{1+c_\nu\epsilon_\nu\tau_{\nu 0}}B_{\nu0}\epsilon_\nu.\]
which leads us to find the same spectrum as ones in Fu11.

For plotting figures (5)-(7), we have normalized the analytical solutions with the surface value of $B_\nu(0)$ and examined several initial optical depths and the photon destruction probability, $\epsilon_\nu$. The parameter $c_\nu$ is set to be $\sqrt{3}$. 

In figure 6, we can see how the mean intensity, $J_\nu$ and the source function, $S_\nu$ change with variation of the optical depth, $\tau_\nu$. Unlike Fig.5, there are two groups of solutions: one with $f_\nu$ (black curves) and the other one with $f_0$ (gray curves). According to panels (a) and (d), $J_\nu$ and $S_\nu$ are apparently descending with respect to $\tau_\nu$ but the physical part of those curves which placed in $\tau_\nu\leq 0$ are approximately constant especially in black colour plots. In this range of the optical depth, i.e. between $\tau_\nu=0$ and $\tau_\nu=0.1$ both constant and variable Eddington factors give equal source function and a bit different mean intensities. In the middle and right-hand side panels with $\tau_{\nu0}\geq 1$, all curves with both assumptions for $f_{Edd}$ display ascending behavior of $J_\nu$ and $S_\nu$ with respect to $\tau_\nu$. Comparing black and gray curves reveals that solutions with $f_\nu$ are smaller than corresponding ones with $f_0$ exception for their initial values (at $\tau_\nu=0$) which are equal.
 
In the uniform heating case, the Planck function varies directly with the optical depth. With the smallest values of the photon destruction probability (means the largest amount of scattering) and the disc optical depth, the differences between $B_\nu$ and other quantities are so remarkable. As $\epsilon_\nu$ increases, the source function and mean intensity become larger and get closer to the Planck function. Therefore, here we have a direct relationship between $\epsilon_\nu$ and radiative quantities including $S_\nu$. From Eq.(34), we find out for $\tau_{\nu0}\rightarrow 0$ we have $H_{\nu0}\sim \epsilon_\nu B_{\nu0}$. Consequently, Eq.(32) leads us to conclude $S_\nu\propto \epsilon_\nu B_{\nu0}$ but just in optically thin discs. According to Eq.(32) and (33), when there is no scattering ($\epsilon_\nu=1$), the source function becomes equal to the Planck function ($S_\nu=B_\nu$). Moreover, we can predict that in the limit of very large disc's optical depth, these three functions $J_\nu$, $S_\nu$ and $B_\nu$ will be equal to:
\[J_\nu(\tau_{\nu0}\rightarrow\infty)= H_{\nu0}\bigg(\frac{\tau_\nu}{f_{Edd}}+c_\nu\bigg)\]
which does not depend on $\epsilon_\nu$'s value, hence we can conclude that the scattering effect is not important in discs with large optical depth.

Using the obtained solutions, we can solve the transfer equation (4) to find the specific intensity. This equation is transformed as an ordinary differential equation: 
\begin{equation} 
\mu\frac{d I_\nu}{d\tau_\nu}- I_\nu =- H_{\nu0}\bigg[ \frac{1}{f_{Edd}}\bigg( 1-\frac{\tau_\nu}{2 \tau_{\nu 0}}\bigg)\tau_\nu+\frac{1+c_\nu\tau_{\nu 0}}{\tau_{\nu 0}}\bigg] 
\end{equation} 
The above equation can be solved analytically to obtain the (outward and inward) intensity, $I_\nu^\pm(\tau_\nu,\mu)$. 
\begin{displaymath} 
I_\nu^\pm(\tau_\nu,\mu)=\mp\frac{H_{\nu0}}{\tau_{\nu0}\mu}\bigg\{ (2\tau_{\nu0}+1) e^{\pm(1 +\tau_\nu)/\mu}Ei\big[\mp\frac{1 +\tau_\nu}{\mu}\big] 
\end{displaymath} 
\begin{displaymath} 
\pm\mu\bigg[\frac{3}{2}\tau_\nu^2-\big[3(\tau_{\nu0}\mp\mu)+1\big]\tau_\nu\pm\mu(3\mu-1) 
\end{displaymath} 
\begin{equation} 
-\tau_{\nu0}(\pm 3\mu+c_\nu-2)\bigg]\bigg\}+C_\pm e^{\pm\tau_\nu/\mu} 
\end{equation} 
as we mentioned before the sign of $\pm$ is positive for outward intensity and it is negative for inward one and $C_\pm\rightarrow C_+$ for outward intensity and $C_\pm\rightarrow C_-$ for inward one. 
Here we also use two proper boundary conditions firstly at the disc's surface ($\tau_\nu=0$) as: 
$$I_\nu^-( 0,\mu )=0,$$ 
and secondly at the disc's midplane ($\tau_\nu=\tau_{\nu0}$) as: 
$$I_\nu^+ (\tau_{\nu 0}, \mu)=I_\nu^-(\tau_{\nu 0},\mu),$$ 
So we can easily find: 
\begin{displaymath} 
C_-=\frac{H_{\nu0}}{\mu\tau_{\nu0}}\bigg\{\mu\bigg[\mu(3\mu+1)+\tau_{\nu0}(3\mu-c_\nu+2)\bigg] 
\end{displaymath} 
\begin{equation} 
-(2\tau_{\nu0}+1)e^{-1/\mu}Ei\big[\frac{1}{\mu}\big]\bigg\} 
\end{equation} 
\begin{figure*} 
\centering 
\includegraphics[width=180mm]{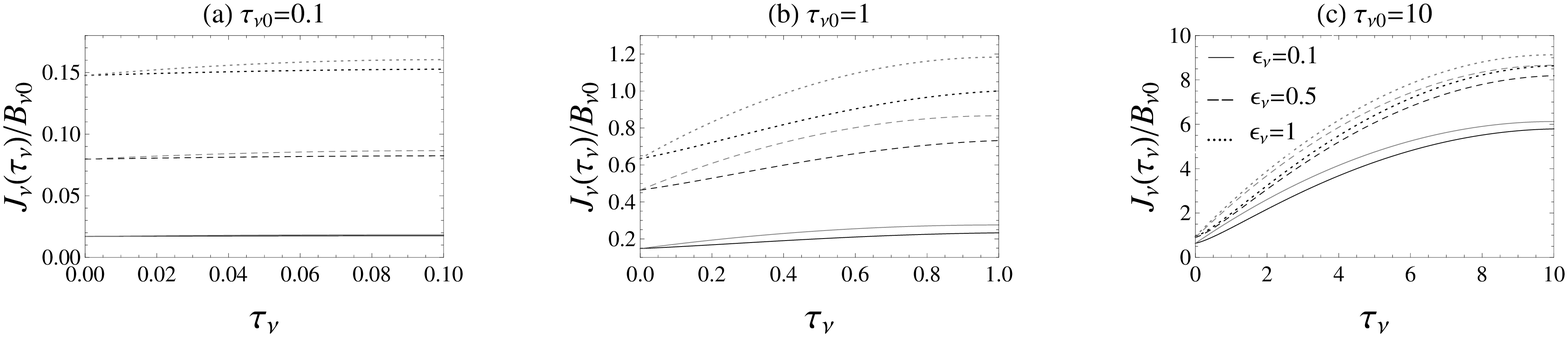} 
\includegraphics[width=180mm]{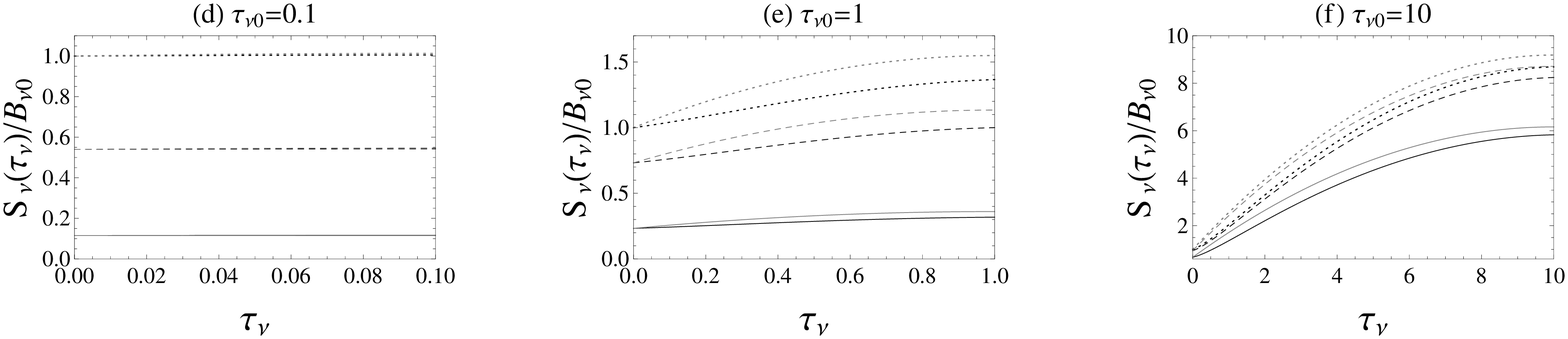} 
\caption{Solutions for uniform heating case: Variations of mean intensity, $J_\nu$, and source function, $S_\nu$ both normalized by $B_{\nu0}$, with respect to the optical depth, $\tau_\nu$. The black curves are referred to solutions obtained by using $f_{Edd}=f_\nu$. In the red curves, the constant Eddington factor, $f_0=\frac{1}{3}$ has been used. } 
\end{figure*} 
\begin{displaymath} 
C_+=\frac{H_{\nu0} e^{-\tau_{\nu0}/\mu}}{\mu\tau_{\nu0}} 
\bigg\{-2\mu^2+(2\tau_{\nu0}+1)\bigg[e^{-(\tau_{\nu0}+1)/\mu }\times 
\end{displaymath} 
\begin{equation} 
Ei\big[\frac{\tau_{\nu0}+1}{\mu}\big]+e^{(\tau_{\nu0}+1)/\mu}Ei\big[-\frac{\tau_{\nu0}+1}{\mu}\big]\bigg\}+C_- e^{-2\tau_{\nu0}/\mu} 
\end{equation} 
The influences of three parameters, $\mu, \epsilon_\nu$ and $\tau_{\nu0}$ on the outward intensity are depicted in figure 8. In the first row panels of Fig.7, we can see for optical depths larger than unity, $I_\nu^+$ with any values of $\epsilon_\nu$ enhances towards the vertical direction. This trend of emergent intensity causes the familiar effect of limb-darkening. Moreover, when the optical depth is unity, $I_\nu^+$ becomes maximum at about $\mu=0.5$. Like Fig.3a, we can see limb-lightening happens for $\tau_{\nu0}=0.1$ and other values smaller than unity and this result is common for all $\epsilon_\nu$'s. According to the second row panels, the photon destruction probability has a direct effect on the emerging radiation. The first row panels also show that the effect of disc optical depth on $I_\nu^+$ is similar to $\epsilon_\nu$'s, i.e. increasing these two parameters makes the intensity rise. As seen, in the uniform heating case, the scattering has a significant effect in our solutions. If we compare gray ($f_{Edd}=f_0$) and black ($f_{Edd}=f_\nu$) curves, we find out the different assumptions for the Eddington factor change remarkably the total shape of each curve with $\tau_{\nu0}\leq 1$, whereas it provides similar shapes. 
\begin{figure*} 
\centering 
\includegraphics[width=180mm]{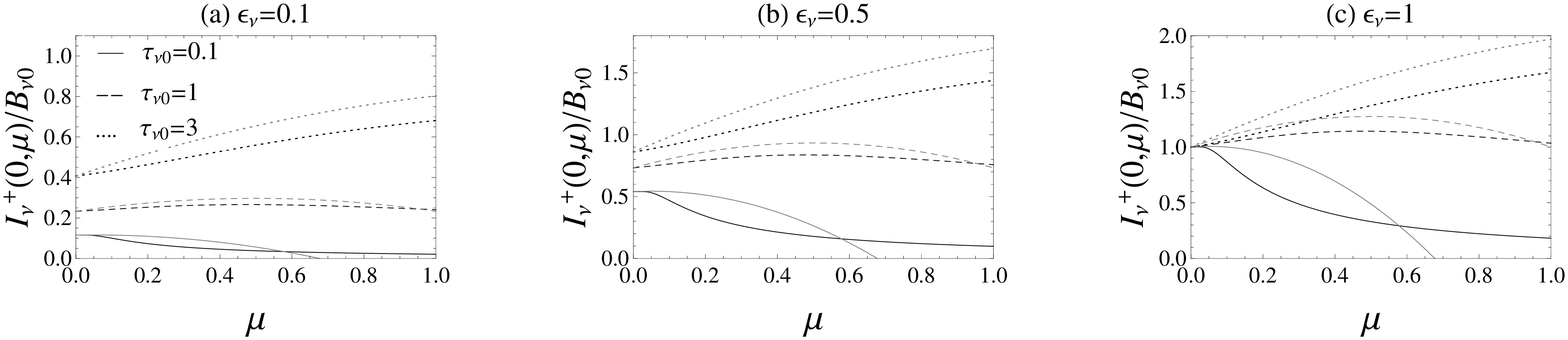} 
\includegraphics[width=180mm]{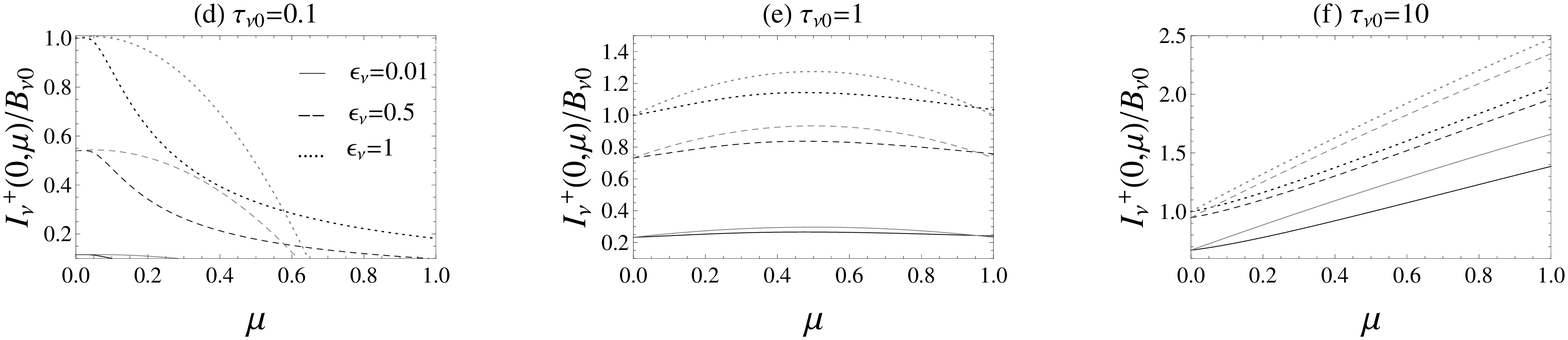} 
\caption{Solutions for uniform heating case: Variation of emergent intensity with respect to the direction cosine, $\mu$. We have examined four values of the photon destruction probability, $\epsilon_\nu$ with the disc optical depth, $\tau_{\nu0}$. The black curves are referred to solutions obtained by using $f_{Edd}=f_\nu$. In the gray curves, the constant Eddington factor, $f_0=\frac{1}{3}$ has been used. Notice that the solutions in panels (a)-(c) are logical for $\tau\leq\tau_{\nu0}$. } 
\end{figure*} 

\section{Frequency dependency of quantities}
Up to now, we have found the functions of radiative quantities  with respect to mainly the optical depth whether the frequency dependency of them were shown implicitly in their indices. In this section, we try to find out how our radiative quantities change with different frequencies. To do that, we need to use dynamical models for knowing the essential parameters of a typical accretion flow. Here, we refer to two models; firstly standard discs of Shakura \& Sunyev (1973) and secondly accretion flows with comparable radiation and gas pressures (AFCRGP) having finite optical depth introduced by Samadi, Abbassi \& Gu (2019).  
  
Before going to these models, we specify the common formula of opacity coefficients.
In high temperature discs with $T\ge 10^4 K$ for pure hydrogen plasmas, the main opacity sources are electron scattering $\sigma_\nu=\kappa_{es}=0.4 cm^2\hspace{0.1cm}g^{-1}$ and free-free absorption:
\begin{displaymath}
\bar{\kappa}=\kappa_{es}+\tilde{\kappa}_{ff},
\end{displaymath}
where free-free absorption, $\kappa_{ff}$ is specified as: 
\begin{equation}
\kappa_{ff}=1.5\times 10^{25}\rho T^{-7/2}\frac{1-e^{-h\nu/k_B T}}{(h\nu/k_B T)^3}\hspace{0.2cm} cm^2\hspace{0.1cm}g^{-1},
\end{equation}
so as seen this quantity depends on temperature, $T$, density $\rho$ and frequency, $\nu$. Nevertheless, to solve dynamical systems, the frequency part of this coefficient is often approximated to a constant value:
\begin{equation}
\tilde{\kappa}_{ff}=6.4\times 10^{22}\rho T^{-7/2}\hspace{0.2cm} cm^2\hspace{0.1cm}g^{-1},
\end{equation}  
so the result of using this formula will be finding a constant photon destruction probability, $\epsilon_\nu=\tilde{\epsilon}$ (independent of frequency) and also a common disc optical depth, $\tau_{\nu0}=\tilde{\tau}_0$ for all photons with any frequency:
\[\tilde{\epsilon}=\frac{\tilde{\kappa}_{ff}}{\tilde{\kappa}_{ff}+\kappa_{es}},\]
\[\tilde{\tau}_0=\int_0^h \rho(\tilde{\kappa}_{ff}+\kappa_{es})dz\]
 Figure 8 displays both forms of these quantities based on input parameters (in Tables 1, 2) of two models, standard and AFCGRP. According to panels (a1, a2), at a certain frquency [$\nu>\nu_m=(10^{15}, 10^{17})$ for $m=(10^8, 10)$, and $\nu>\nu_m=10^{16}$ in the second model], the total optical depth, $\tau_{\nu0}$ tends to a constant value, equals $\tilde{\tau}_0$ (gray horizontal lines). In the second column panels, we see $\epsilon_\nu$ is unity for frequencies smaller than $\nu_m$. Moreover, the opacity coefficient of free-free absorption is larger than the Thomson scattering at the range of $\nu<\nu_m$ and becomes almost ignorable with frequencies about ten times larger than $\nu_m$. 
   
\begin{figure*} 
\centering 
\includegraphics[width=180mm]{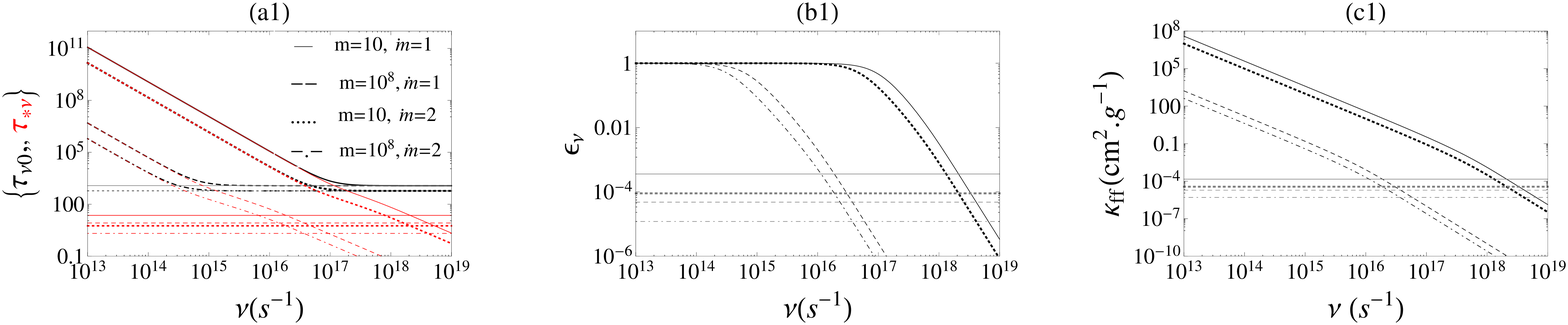} 
\includegraphics[width=180mm]{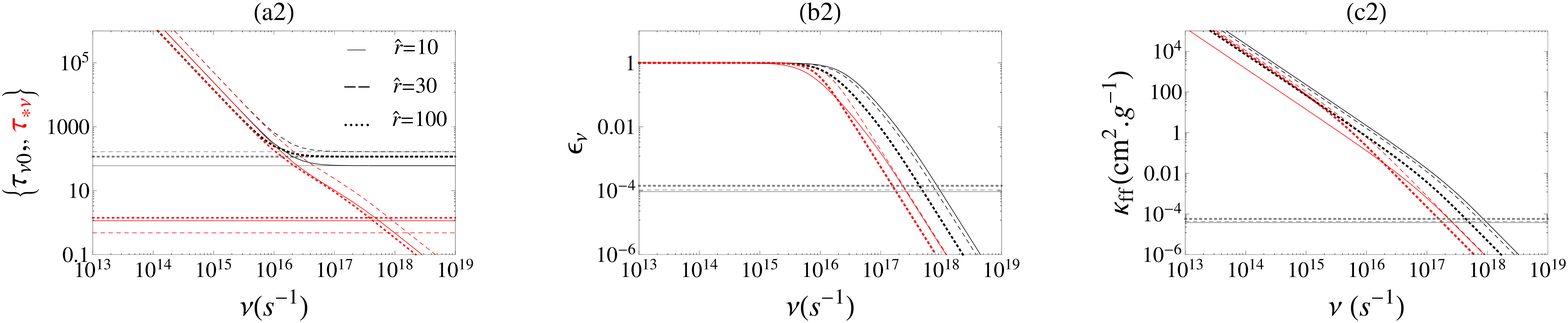} 
\caption{Variations of  (a) total and effective optical depth of disc, $\tau_{\nu0}, \tau_{*\nu}$  (b) the photon destruction probability, $\epsilon_\nu$,   (c) free-free absorption opacity coefficient, $\kappa_{ff}$, with respect to frequency, $\nu$.
The first row panels display the radiative parameters of four standard discs at $r=4r_g$ with $\alpha=1$, $T=T_c$ and $\rho=\rho_c$ (where $T_c$ and $\rho_c$ are the temperature and density at the disc's equator, see Table 1). In the panels of second row, the input  parameters from Table.2 have been used to show absorption properties of an accretion flow with comparable gas and radiation pressure (AFCGRP) at three different radii: $r=(10,30,100)r_g$.  In this figure, the horizontal lines show the approximated value of each quantity (see $\tilde{\kappa}_{ff}, \tilde{\epsilon}, \tilde{\tau}_0, \tau_*$ in Table.1, 2). The red plots in panels (b2) and (c2) show the surface values of $\epsilon_\nu$ and $\kappa_{ff}$ (we have used $\rho_s$ and $T_s$ of Table.2 in Eq.39 ).  } 
\end{figure*} 

\subsection{Standard Discs}
Shakura-Sunyev discs are well-known as geometrically thin but optically thick. At the inner region of standard discs, we find density, temperature, scale height and effective opacity as certain functions of these nondimensional quantities: 1. the central mass, $m=M/M_\odot$, 2. the mass accretion rate, $\dot{m}=\dot{M}/\dot{M}_{crit}$ (where $\dot{M}_{crit}=2.22\times 10^{-9} m M_\odot yr^{-1}$) 3. the viscosity parameter, $\alpha$ and 4. the radius of observation, $\hat{r}=r/r_g$ (where $r_g=2 GM/c^2$):
\begin{displaymath}
\rho=9.0\times 10^{-4} (\alpha m)^{-1} \dot{m}^{-2}\hat{r}^{3/2}f^{-2} \hspace{0.3cm}g \hspace{0.1cm} cm^{-3}
\end{displaymath}
\begin{displaymath}
T_c=4.9\times 10^7(\alpha m)^{-1/4}\hat{r}^{-3/8}\hspace{0.3cm} K
\end{displaymath}
\begin{displaymath}
H=5.5\times 10^4  m \dot{m} f\hspace{0.3cm} cm
\end{displaymath}
where $f=1-\sqrt{3r_g/r}$ (Kato et al. 2008). Beside the total optical depth of disc, we define the effective optical depth as $\tau_*=\sqrt{\bar{\kappa}\tilde{\kappa_{ff}}}\rho H$:
\begin{displaymath}
\tau_*= 8.4\times 10^{-3}\alpha^{-17/16} m^{-1/16}\dot{m}^{-2}\hat{r}^{93/32} f^{-2},
\end{displaymath} 
At the inner region of standard disc, the total optical depth of disc ($\tilde{\tau}_0=\bar{\kappa}\rho H$) becomes:
\begin{displaymath}
\tilde{\tau}_0=2.0\times 10^1 (\alpha\dot{m})^{-1}\hat{r}^{3/2}f^{-1},
\end{displaymath}
Now, we can calculate all these quantities for a set of input parameters: $(m,\dot{m},r,\alpha)$. The results for $\hat{r}=4, \alpha=1, m=10, 10^8$ and $\dot{m}=1, 2$ are listed in Table 1. 
In panel (a1) of figure 8, the red plots show $\tau_{*\nu}=\sqrt{\bar{\kappa}\kappa_{ff}}\rho H$ which is equal to $\tau_{\nu0}$ (black plots) at $\nu\leq\nu_m$.  

\begin{table} 
\caption{Quantities at the inner region of four standard discs with $\alpha=1$ at $r=4r_g$.} 
\centering 
\begin{tabular}{c rrrr} 
\hline 
($m, \dot{m}$) & (10, 1) & ($10^8$, 1) & (10, 2) & ($10^8$, 2) \\ 
\hline 
$T_c/10^7 (K)$ & 1.64 & 0.03 & 1.64 & 0.03 \\ 
$(H/r)/10^{-2}$& 0.62 & 0.62 & 1.24 & 1.24 \\
$\Sigma/10^3 (g.cm^{-2})$ & 5.91 & 5.91 & 2.95 &2.95\\
$\rho_c (g.cm^{-3})$ & 0.04 & $\frac{4.01}{10^9}$& 0.01 & $\frac{1.00}{10^9}$ \\ 
$\tilde{\kappa}_{ff}/10^{-4}$ ($cm^2.g^{-1}$)  & 1.44 & 0.19 & 0.36 & 0.05\\ 
$\tilde{\epsilon}/10^{-4}$ & 3.60 & 0.48 & 0.90 & 0.12\\ 
$\tilde{\tau}_0/10^3$ & 1.19 & 1.19 & 0.59 & 0.59 \\ 
$\tau_{*}/10$ & 2.27 & 0.83 & 0.57 & 0.21 \\ 
[1ex] 
\hline 
\end{tabular} 
\label{tab:hresult} 
\end{table} 
Some basic assumptions (or approximations) have been used to obtain formulas of standard discs: the radiation pressure is dominated at the inner region, $p\sim p_{rad}$, and also the main opacity is due to electron scattering, $\bar{\kappa}\sim \sigma_{es}$ which is confirmed in Table 1 ($\tilde{\kappa}_{ff}\rightarrow 0$). In panel (c1) of figure 8, we see clearly that the valid frequency range for this assumption is $\nu\gtrsim 10\nu_m$.

In figure 9, we have presented the frequency dependency of the source and Planck functions for two cases of radiative equilibrium (RE) and uniform heating (UH). As we mentioned before, $S_\nu, B_\nu$ and $J_\nu$ are equal in RE case, so panels (a) and (b) show all these three functions. As seen, the difference between source functions with constant ($f_0$) and variable ($f_\nu$) Eddington factor is larger at the surface of disc in comparison with $S_\nu$ at photosphere (i.e. $\tau_\nu=1$). The black solid curves in Fig.9a,b show the simple form of Planck function, $B(\nu,T_c)=2h\nu^3 c^{-2}[Exp(h\nu/k_BT)-1]^{-1}$, multiplied by frequency, $\nu$ which is equal to $\nu S_\nu(\tau_\nu=0,f_0)$ for RE case and also equals $B_\nu$ in Eq.(33) for both mass accretion rates $\dot{m}=1,2$ and both $f_0, f_\nu$ for UH case. Panels (c) and (d) represent the photosphere's radiations from two systems with $m=10, 10^8$, and $\dot{m}=1, 2$, based on formulas of UH case. Like differences of $S_\nu$'s with $f_0$ and $f_\nu$ at the surface, the source function here at the photosphere is smaller with $f_\nu$ than with $f_0$.  The dotted and dot-dashed plots are Planck functions (blue ones with $f_\nu$ and dot-dashed black one with $f_0$).  As seen, $B_\nu$ with $\dot{m}=1$ is greater than one with $\dot{m}=2$. 

 \begin{figure*} 
\centering 
\includegraphics[width=165mm]{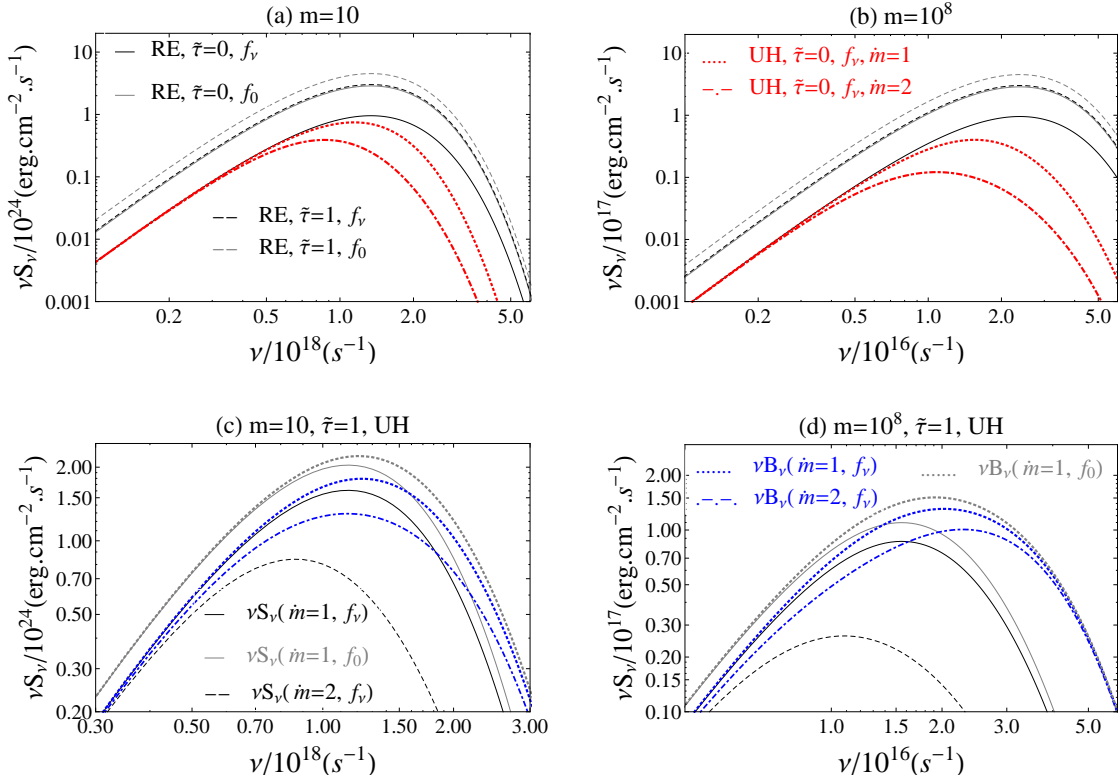} 
\caption{The frequency dependency of source function related to standard discs for two cases of radiative equilibrium (RE) and uniform heating (UH). In this figure, we have examined two different optical depths (surface $\tilde{\tau}=0$ and photosphere $\tilde{\tau}=1$), central masses ($m=10,10^8$) and mass accretion rates ($\dot{m}=1,2$). The gray plots show functions with $f_{Edd}=f_0$. Planck function at photosphere is different from source function in UH and seen in blue colour in panels (c) and (d).    } 
\end{figure*} 
\subsection{Accretion flows with comparable radiation and gas pressures}
In standard discs, we had analytical solutions to calculate temperature and density and their very large total optical depth. Here, we want to use another model a bit different for finding the spectrum of an accretion system with less optical depth. Unlike the inner region of standard discs, here we have gas pressure, $p_g$ beside radiation pressure, $p_r$ which are comparable. We define $\beta$ parameter as the ratio of $p_g$ to the total pressure, $p_t=p_g+p_r$ (so $\beta=p_g/p_t$). This parameter varies with vertical position and it is specified by $\beta_c$ at the equatorial plane ($c$ index means the value of quantity at the disc's equator). The main difference of this model with standard disc is that two separate energy equations for matter and radiation in the diffusion limit are considered (see Eq.4, 5 of Samadi et al. 2019) and the self-similar technique in the radial direction has been employed (for instance density changes as $\rho\propto r^{-3/2}$). Furthermore, in addition to the radiation cooling, some percentage energy ($f_{adv}=Q_{adv}/Q_{vis}$) of viscous heating is transported in the radial direction and advected towards the central object. As an input parameters, we choose $\beta_c=1, \alpha=0.1, \gamma=1.5, Z=0$ (where $Z$ is the metallicity, so we have assumed that bound-free absorption does not happend) and $\rho_c= 4\times 10^{-4}$ at $r=30r_g$. The other quantities are found from numerical solutions and listed in Table 2. In this table, $\rho_1, T_1$ and $\tilde{\epsilon}_1$ have been determined at $\tilde{\tau}=1$. As seen, $\tilde{\tau}_0$ and $\tilde{\tau}_*$ are one order of magnitude smaller than standard disc's total and effective optical depths.

Knowing  the surface values of temperature ($T_s$) and density ($\rho_s$) enables us to evaluate radiation quantities exactly at the surface.  In panels (b2) and (c2) of Fig.8, the red plots have been produced by substituting $T_s, \rho_s$ in Eq.(39) and (40), whereas the black ones are founded with using $T_c, \rho_c$ (the equatorial values) in those two equations. On the other hand, for AFCGRP model, we have more options of input parameters to produce the spectrum especially for UH case as seen in figure 10. In panel (a), we have plotted the spectrum from the three radii: $r=(10,30,100)r_g$ by employing formulas of $S_{\nu}$ (or $J_{\nu}$ in Eq.18) with $f_{\nu}$ and using $T_c$ and $T_s$ for each radius (see table 2). The photosphere's radition for RE case is illustrated in panel (b). In order to compare solutions with two values of $f_{Edd}$, we have brought $S_\nu$'s with $f_0$ at $\tilde{\tau}=0, 1$ for the smallest radius, $\hat{r}=10$ in panel (b). As seen, the difference between $S_{\nu}$'s with $f_0$ and $f_{\nu}$ is larger at the surface (gray and black plots) in comparison with them at photosphere (blue and red plots). The interesting point of this panel is that two (red and black) graphs coincide which implies the radiation from photosphere with $f_\nu$ and from the surface but with the equatorial temperature, $T_c$ are approximately equal. In the plots of $r=10r_g$ in Fig.10c for UH case involve $\epsilon_{\nu}, \tilde{\epsilon}$ and $\tau_{\nu0}, \tilde{\tau}_0$, hence we have two $S_{\nu}$'s  with noticeable differences for each optical depth. Moreover, like RE case, functions with constant and variable Eddington factor have approximately the same result at the disc's photosphere.       
  
\begin{table} 
\caption{Quantities of an accretion flow with comparable radiation and gas pressures, $\alpha=0.1, \beta_c=0.1, \gamma=1.5, m=10$. } 
\centering 
\begin{tabular}{c rrr} 
\hline 
$r/r_g$& 10 & 30 & 100 \\
\hline 
$\rho_c/10^{-4}(g.cm^{-3})$ & 7.348  & 4.000 &0.657 \\ 
$\rho_1/10^{-5}(g.cm^{-3})$ & 4.443  & 0.978 &0.196\\ 
$\rho_s/10^{-6}(g.cm^{-3})$ & 7.973  & 3.209 &0.566 \\ 
$T_c/10^6(K)$&7.627 &6.195& 3.391\\
$T_1/10^6(K)$&7.503 &5.782& 3.229\\ 
$T_s/10^6(K)$&2.398 &0.331& 0.122\\ 
$\dot{m}/10^{-1}$ &0.271 &3.077& 7.490\\ 
$(H/r)/10^{-2}$ &1.720 &2.960 & 3.919\\
$\Sigma/10^2 (g.cm^{-2})$&3.031 & 8.310&5.820\\
$f_{adv}/10^{-2}$ &0.343 &1.108&2.069\\
$\tilde{\kappa}_{ff}/10^{-5}$&3.743&4.217 &5.711\\ 
$\tilde{\epsilon}_c/10^{-4}$&0.936&1.054 &1.427\\ 
$\tilde{\epsilon}_1/10^{-6}$&5.990&3.284 &5.058\\ 
$\tilde{\epsilon}_s$&$\frac{5.820}{10^5}$&0.997 &0.121\\ 
$\tilde{\tau}_0/10^2$ & 0.606 &1.662 & 1.164   \\ 
$\tau_{*}$ & 0.483 & 1.422 & 1.148 \\ 
[1ex] 
\hline 
\end{tabular} 
\label{tab:hresult} 
\end{table} 
\begin{figure*} 
\centering 
\includegraphics[width=160mm]{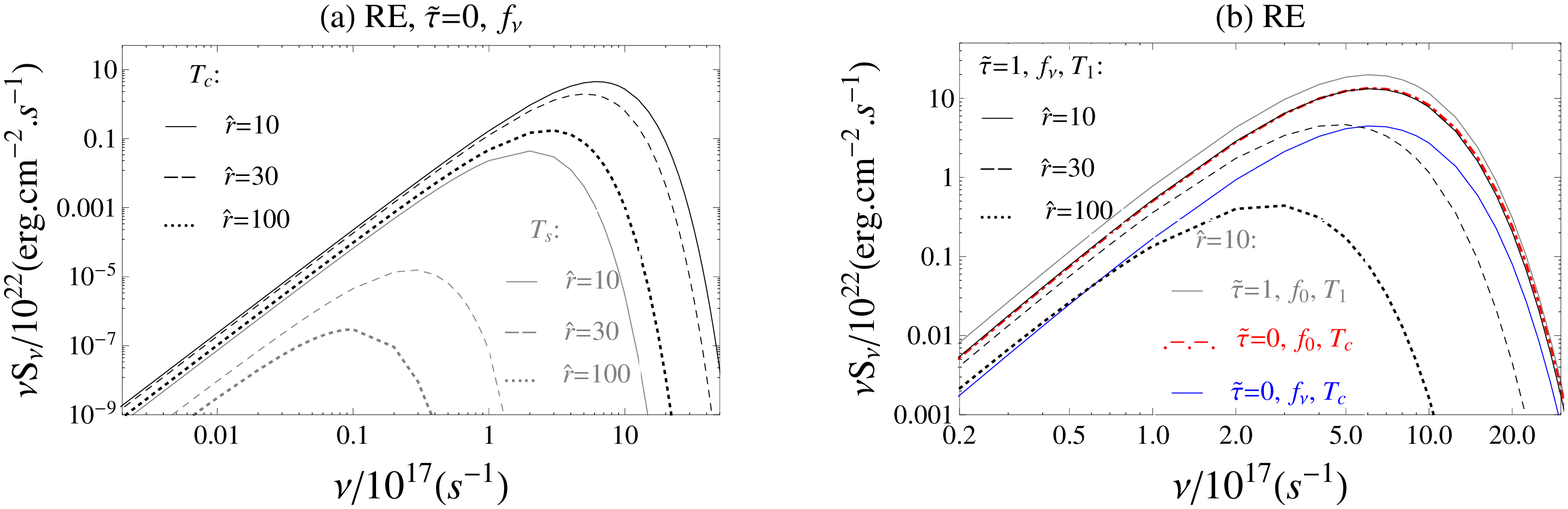} 
\includegraphics[width=120mm]{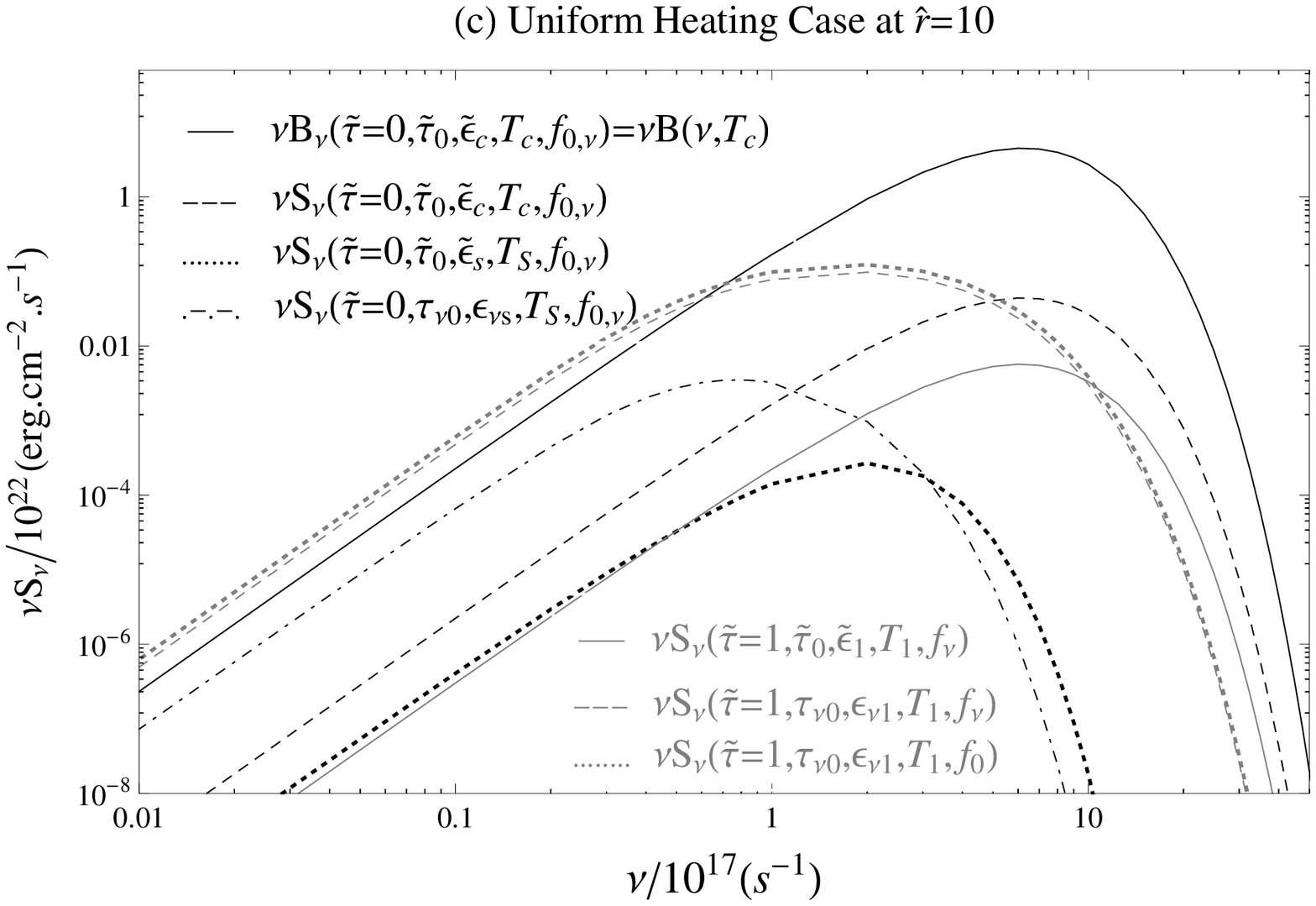} 
\caption{The frequency dependency of source function, $S_\nu$, and Planck function, $B_\nu$, produced by the second model (AFCGRP)  originates from the surface ($\tilde{\tau}=0$) and photosphere ($\tilde{\tau}=1$) of disc. 
 In two upper panels, the spectrum is based on RE case and from three different radii, $r=(10,30,100)r_g$. In panel (a), the surface source function is calculated with using temperature of the equator, $T_c$, (black plots) and surface, $T_s$ (gray ones). In panel (b), four plots of $S_\nu$ are related to photosphere with $\tilde{\tau}=1, T_1$ (black and gray plots in panel b) and the two other plots show the surface radiation with $f_0$ (red dot-dashed) and $f_\nu$ (blue solid). Notice all applied temperatures here are listed in table 2. In panel (c), for calculating $S_\nu$ we have used three different  photon destruction probability parameter: three ones for the surface, $\tilde{\epsilon}_s, \epsilon_{\nu s}, \tilde{\epsilon}_c$, two ones for the photosphere, $\tilde{\epsilon}_1, \epsilon_{\nu 1}$ and we have also employed two total optical depths: $\tau_{\nu0}, \tilde{\tau}_0$ ($\nu$ index indicates the frequency dependency and related to Eq.39). } 
\end{figure*}
\section{Summary and Conclusions} 
In this work, we focused our attention to solve analytically radiative transfer equations related to a geometrically thin accretion disc with a finite optical depth. We simplified the basic equations by using plane-parallel approximation and also other several assumptions which helped us to solve analytically this problem. We considered three different cases: (i) radiative equilibrium (RE), and (iii) a flow with uniform internal heating. Moreover, both cases were supposed to be in local thermodynamic equilibrium (LTE). We employed Eddington approximation to have access a relationship between two radiative quantities, $J_\nu$ (mean intensity) and $K_\nu$ (mean radiative stress). To achieve more accurate solutions, we took into account the variable Eddington factor, $f_{Edd}=f_\nu=(1+\tau_\nu)/(1+3\tau_\nu)$. We compared our results with those obtained by constant Eddington factor, i.e. $f_0=1/3$ which has been used in Fu11. 

We also studied the dependency of solutions to these main parameters: the optical depth, $\tau_\nu$, the direction cosine, $\mu(=\cos\theta)$, the total disc optical depth, $\tau_{\nu0}$, the photon destruction probability, $\epsilon_\nu$ (which has an opposite relation with scattering, i.e. scattering is maximum when $\epsilon_\nu$ is zero). 
For the RE case ($j_\nu=4\pi\kappa_\nu J_\nu$), we found a constant Eddington flux ($H_\nu$) and three equal linear functions of the optical depth, consisting of the mean intensity, $J_\nu$, the source function, $S_\nu$, and the Planck function, $B_\nu$ (for LTE $B_\nu$=$J_\nu$). To achieve the specific intensity, $I_\nu$, we solved the differential equation of radiative transfer with respect to one explicit variable of the optical depth. If we employed $f_0$, it would give a relatively simple relation for $I_\nu$ including a linear term and an exponential function with respect to $\tau_\nu$. With $f_\nu$, a more complicated function was found for the specific intensity including an exponential integral function beside the linear term. The boundary conditions were needed to complete the solutions. One of them was found by ignoring irradiation and using of null incident intensity at the disc's surface (where $\tau_\nu=0$). The other boundary condition was related to the absence of equatorial heating and based on it we applied the balance between outward ($I_\nu^+$) and inward ($I_\nu^-$) intensities at the equatorial plane (where $\tau_\nu=\tau_{\nu0}$). Comparing our solutions with constant and variable Eddington factors, we noticed the main difference in their values but similarity in their total trend. In a disc with small optical depth, the emergent intensity became independent of its optical depth and hence it looked like an optically thick disc. For RE case, we found out that scattering is not effective and solutions are independent of $\epsilon_\nu$. For the other case, this factor appeared so important and caused significantly changes in all plots. 

In the second case, we concentrated on discs with uniform internal heating and achieved analytical solutions with $f_\nu$. 
The plots revealed that the mean intensity and Eddington flux are more sensitive to the scattering factor in optically thin discs, but the plots of these two quantities with respect to the optical depth illustrated they grows in the opposite directions. Unlike $H_\nu$ and $J_\nu$, we encountered dissimilar trends of the mean radiative stress, $K_\nu$, with respect to $\tau_\nu$ under the effect of $\tau_{\nu0}$. Regarding radiation from the disc's surface, we found out that the parameter of $\epsilon_\nu$ has a positive effect on growing the emergent intensity $I_\nu^+(\tau_\nu=0,\mu)$ especially in discs with smaller optical depth. The other point was that in optically thin discs with $\tau_{\nu0}<1$, we saw limb-brightening instead of limb-darkening which was a common result for both cases in this work. 

We also studied the frequency dependency of the radiative quantities named in this paper. We calculated temperature and density based on two dynamical models of accretion systems and employed them in plotting spectrum from zero and unity optical depth. The two relationships for the free-free absorption coefficient made more than one curve for the spectrum. Moreover, knowing the vertical dynamical structure of the second model enabled us to find out more points and details about the radiation profile of such a system.

Although we tried to achieve more precise results in this paper, we had to use a lot of implicit and explicit simplifying assumptions such as: 1. solving the set of radiative equations separately from the dynamical part, 2. considering just coherent electron scattering and neglecting other kinds of scattering, 3. ignoring line opacity effects, 4. excluding convection, conduction and irradiation 5. neglecting any movements in the flow, static atmosphere was considered, ... . It is still a very long way to improve our theoretical study by using less assumptions and find more real solutions which fit the data of observed accretion disc's spectra. 

\section*{Acknowledgements}
We are grateful to the anonymous referee for his/her thoughtful and
constructive comments which enabled us to improve the first edition
of this paper. This work has been supported financially by Research
Institute for Astronomy and Astrophysics of Maragha (RIAAM)
under research project No. 1/62753.

\end{document}